\documentclass[%
 reprint,
superscriptaddress,
nolongbibliography,
 amsmath,amssymb,
 aps,
floatfix,
]{revtex4-2}

\usepackage{color}
\usepackage{graphicx}
\usepackage{dcolumn}
\usepackage{bm}
\usepackage[colorlinks= true,urlcolor=blue,linkcolor= blue,citecolor=blue,bookmarks=false,pdfstartview=]{hyperref}
\usepackage{color}

\begin{document}

\preprint{APS/123-QED}

\title{Spin dynamics in the Dirac $U(1)$ spin liquid YbZn$_2$GaO$_5$}
\author{Hank C. H. Wu}
 \email{hank.wu@physics.ox.ac.uk}
\affiliation{Clarendon Laboratory, Department of Physics, University of Oxford, Parks Road, OX1 3PU, United Kingdom}%
\author{Francis L. Pratt}
\affiliation{ISIS Facility, Rutherford Appleton Laboratory, Chilton, Oxfordshire OX11 0QX, United Kingdom}%
\author{Benjamin M. Huddart}
\affiliation{Clarendon Laboratory, Department of Physics, University of Oxford, Parks Road, OX1 3PU, United Kingdom}%
\author{Dipranjan Chatterjee}
\affiliation{Clarendon Laboratory, Department of Physics, University of Oxford, Parks Road, OX1 3PU, United Kingdom}%
\author{Paul A. Goddard}
\affiliation{Department of Physics, University of Warwick, Coventry CV4 7AL, United Kingdom}
\author{John Singleton}
\affiliation{National High Magnetic Field Laboratory, Los Alamos National Laboratory, Los Alamos, New Mexico 87545, USA}
\author{D. Prabhakaran}
\affiliation{Clarendon Laboratory, Department of Physics, University of Oxford, Parks Road, OX1 3PU, United Kingdom}%
\author{Stephen J. Blundell}
 \email{stephen.blundell@physics.ox.ac.uk}
\affiliation{Clarendon Laboratory, Department of Physics, University of Oxford, Parks Road, OX1 3PU, United Kingdom}%

\date{\today}%

\begin{abstract}
YbZn$_2$GaO$_5$ is a promising candidate for realizing a quantum spin liquid (QSL) state, particularly owing to its lack of significant site disorder. Pulsed-field magnetometry at 0.5 K shows magnetization saturating near 15 T, with a corrected saturation moment of 2.1(1) $\mu_\mathrm{B}$ after subtracting the van Vleck contribution.  Our zero-field $\mu$SR measurements down to milliKelvin temperatures provide evidence for a dynamic ground state and the absence of magnetic order.  To probe fluctuations in the local magnetic field at the muon site, we performed longitudinal field $\mu$SR experiments.  These results provide evidence for spin dynamics with a field dependence that is consistent with a U1A01 Dirac QSL as a plausible description of the ground state. 
\end{abstract}

\maketitle

Frustrated magnetic materials with competing interactions have proven to be
a fertile arena for the discovery of novel emergent phenomena \cite{Lacroix2011,Starykh2015}. In
these systems, frustration disrupts the development of a conventional
long-range ordered ground state and can result in the formation of a quantum spin liquid (QSL), a state which is characterized by a high degree of quantum entanglement and topological order \cite{Wen2017} and which often hosts novel, fractionalized
emergent excitations \cite{Zhou2017,Balents2010,Savary2017,Lancaster2023}.  
There are many different theoretical
models describing different types of QSL, such as gapless $Z_2$ fermions on the Kitaev honeycomb lattice \cite{Kitaev2006,Trebst2022}, gapped $Z_2$ topological states on the kagome lattice \cite{Moessner2001,Depenbrock2012}, and gapless Dirac fermions with emergent $U(1)$ gauge fields~\cite{Ran2007,Iqbal2013,Hastings2000}.
Many exotic cases of geometrically
frustrated quantum magnets are found in two-dimensional systems, with
particular interest in materials that host triangular lattices~\cite{Zhu2018,Zheng2024_MagOsc,Zheng2024_Plateau}, thus motivating the
design and synthesis of new materials with perfect triangular lattices \cite{Norman2016}.
Rare-earth-containing compounds can be attractive QSL candidate
materials \cite{Liu2018}.  In these systems the subtle interplay between frustration
due to the geometry of the lattice, the spin-orbit coupling, and
the crystal field anisotropy of the rare earth ion, can lead to a
variety of rich emergent phenomena \cite{Broholm2020,Clark2021}. 
Yb$^{3+}$ (4f$^{13}$) is an important rare earth ion in this context, as its ground-state Kramers
doublet, which results from crystal-field splitting of the eightfold degeneracy of its $J=\frac{7}{2}$ moment behaves as an effective spin-$\frac{1}{2}$ state.  Interest has therefore focused on YbMgGaO$_4$~\cite{Li2015,Li2016} which contains Yb$^{3+}$ ions arranged in a perfect triangular lattice. However, chemical disorder due to mixed
Mg and Ga occupancies complicates the interpretation of data from this compound since
disorder can mimic the continuum-like inelastic
neutron scattering (INS)
signal that would otherwise be interpreted as resulting from spinon
excitations of a QSL state \cite{Zhenyue2017,Paddison2017};
a similar issue affects the sister compound YbZnGaO$_4$~\cite{Pratt2022}.  For this reason, there
is a very strong motivation to study the newly discovered related material
YbZn$_2$GaO$_5$ which (like YbMgGaO$_4$ and YbZnGaO$_4$)
hosts an ideal triangular lattice of
effective spin-$\frac{1}{2}$ moments
but crucially (and unlike YbMgGaO$_4$ and YbZnGaO$_4$)
has no inherent chemical disorder~\cite{Bag2024}.

\begin{figure}[hb]
\includegraphics[width = 0.47\textwidth]{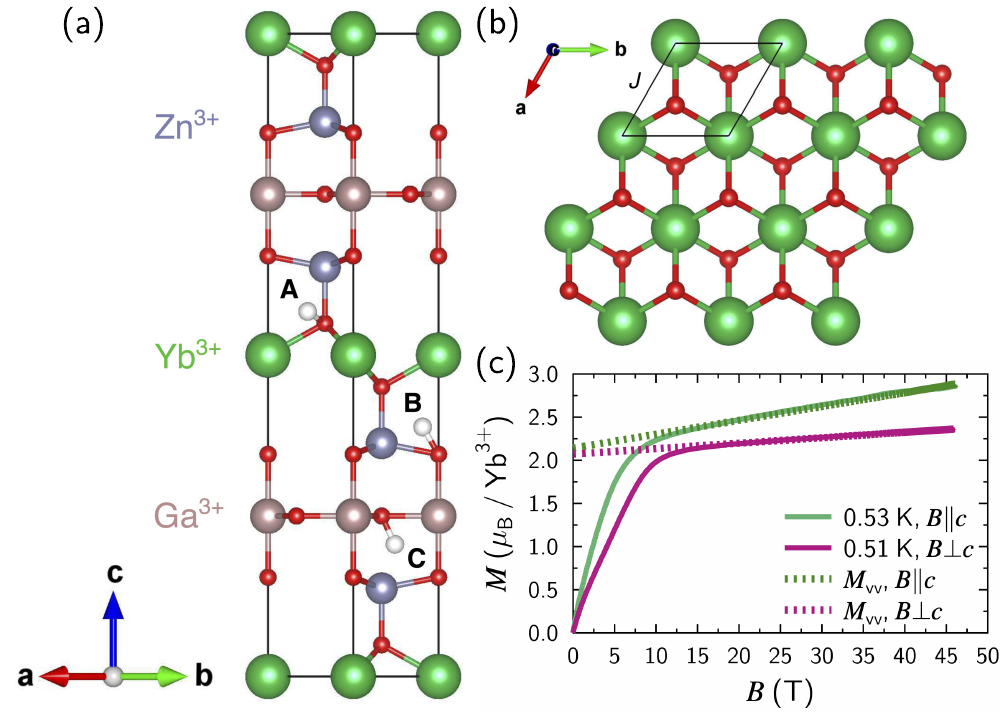}
\caption{\footnotesize 
(a) Crystal structure of YbZn$_2$GaO$_5$ viewed along the $b$ axis; Yb-O planes are
  well separated by non-magnetic Zn-O, Ga-O, and Zn-O layers along the
  crystallographic $c$ axis. The muon sites A, B and C are marked, each bonded to a distinct oxygen atoms (red spheres) in the structure.
(b) The Yb$^{3+}$ ions (green spheres)
  form a triangular lattice with nearest neighbor exchange coupling $J$.
(c) Pulsed-field magnetization for fields parallel (green) and perpendicular (violet) to the crystallographic $\mathbf{\hat{c}}$ axis, with the dashed lines accounting for the linear van Vleck contributions.
}
\label{fig1}
\end{figure}

The crystal structure of YbZn$_2$GaO$_5$ is shown in Fig.~\ref{fig1}(a), with
the well separated triangular layers illustrated in Fig.~\ref{fig1}(b).
Magnetic susceptibility ($\chi$) data~\cite{Bag2024} show no evidence of
magnetic ordering down to 0.3~K, while INS data~\cite{Bag2024} reveal a
broad continuum of scattering which remains gapless between the M and
K points, but is gapped near the $\Gamma$ point; these properties are
consistent with the presence of a $U(1)$ Dirac QSL state~\cite{Moore2023}. This is a particular type of gapless QSL that
hosts fermionic excitations that couple to a $U(1)$ gauge field (as in conventional electromagnetism) and contains Dirac cones in the spinon excitation spectrum.
We prepared single crystal samples of YbZn$_2$GaO$_5$ using the previously established method~\cite{Bag2024} and characterized our sample using x-ray diffraction and SQUID magnetometry.
We also used pulsed-field magnetometry at 0.5~K to extend these data to fields of up to 45~T, measuring the magnetization of the YbZn$_2$GaO$_5$ sample parallel and perpendicular to the crystallographic $c$-axis. In both directions, the magnetization begins to saturate at around 15~T above which field only the linear van Vleck magnetization contributes. By subtracting this van Vleck contribution, both sets of measurements extrapolate back to a saturation moment of $2.1(1)~\mu_\text{B}$, as shown in Fig.~\ref{fig1}(c).

\begin{figure}[htb]
\includegraphics[width =0.49\textwidth]{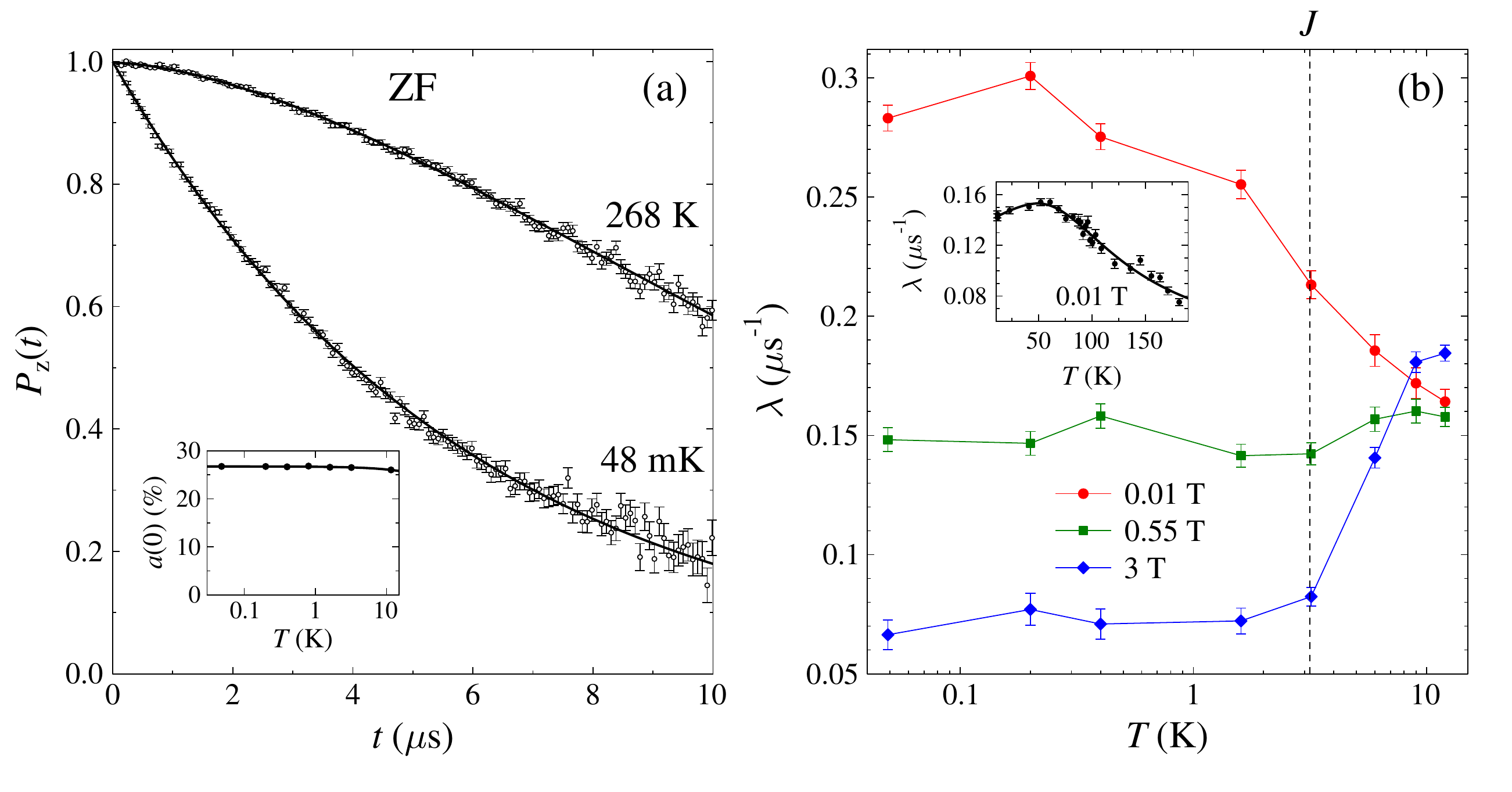}
\caption{\footnotesize 
(a) Zero field relaxation at 268~K and 48~mK.
The inset shows that there is no significant change in the initial asymmetry at low $T$.
(b) LF relaxation rate versus $T$ in the low-$T$ region.
The dashed line shows the energy of the exchange coupling estimated from magnetic susceptibility measurements.
The inset shows the relaxation rate in 0.01~T extended to the high-$T$ region where thermal excitation of crystal field levels takes place. The solid line is a model fit used to estimate the crystal field excitation energy. 
}
\label{fig2}
\end{figure}

We used muon spin rotation ($\mu$SR) measurements
to extract the spin dynamics in
YbZn$_2$GaO$_5$, since muons are an extremely effective probe of slow spin fluctuations \cite{Pratt2023,Blundell2022}.
Our measurements were performed using the HiFi $\mu$SR instrument at the ISIS Neutron and Muon Source and used both zero-field (ZF) and longitudinal-field (LF) configurations.
The sample was a mosaic of crystals aligned so that the crystallographic $c$ axis is parallel to the muon beam and to the applied LF. Sample cooling was performed using a helium dilution refrigerator providing temperatures down to 48~mK. Data analysis was performed using the WiMDA program~\cite{Pratt2000}.

The measured forward to backward asymmetry of the muon decay positrons $a(t)$ was fitted to a single relaxation component plus a constant background, consistent with a single dominant muon site, i.e.\
$a(t) = a_\mathrm{0} P_z(t) + a_\mathrm{BG}$, where $P_z(t) = \exp\left(-({\lambda}t )^\beta\right)$ is a stretched exponential function and $\lambda$ is the muon spin relaxation rate. 
In the low-temperature region, the measured ZF relaxation was well described by a simple exponential ($\beta$=1, signifying spin dynamics with a single fluctuation rate) and no oscillations were observed at the base temperature 48~mK, as shown in Fig.~\ref{fig2}(a). This indicates an absence of magnetic order and a dynamic ground state, which would be consistent with QSL behavior.

To probe the nature of the spin fluctuations, we studied the effect of an applied LF since such measurements can reveal the spectral density of the spin fluctuations \cite{Blundell2022}. The measured dependence of the relaxation rate $\lambda$ on field and temperature is given in Fig.~\ref{fig2}(b).  We find strong field dependence in $\lambda$ at low temperatures and much weaker field dependence at 5~K and above. 
The low-field relaxation has been followed to higher temperatures to explore crystal field excitations with the relaxation rate being fitted to an activation dependence $\lambda^{-1} = \nu_0 + \nu_1 \exp(-E_\mathrm{A}/k_{\rm B}T)$, where $\nu_0$ is given a weak linear $T$ dependence for the best fit [Fig.~\ref{fig2}(b), inset] and we obtain $E_\mathrm{A}$~=~25(2)~meV, broadly consistent with the calculated crystal field energy levels (see End Matter).

\begin{figure*}[htb]
\includegraphics[width = 0.9\textwidth]{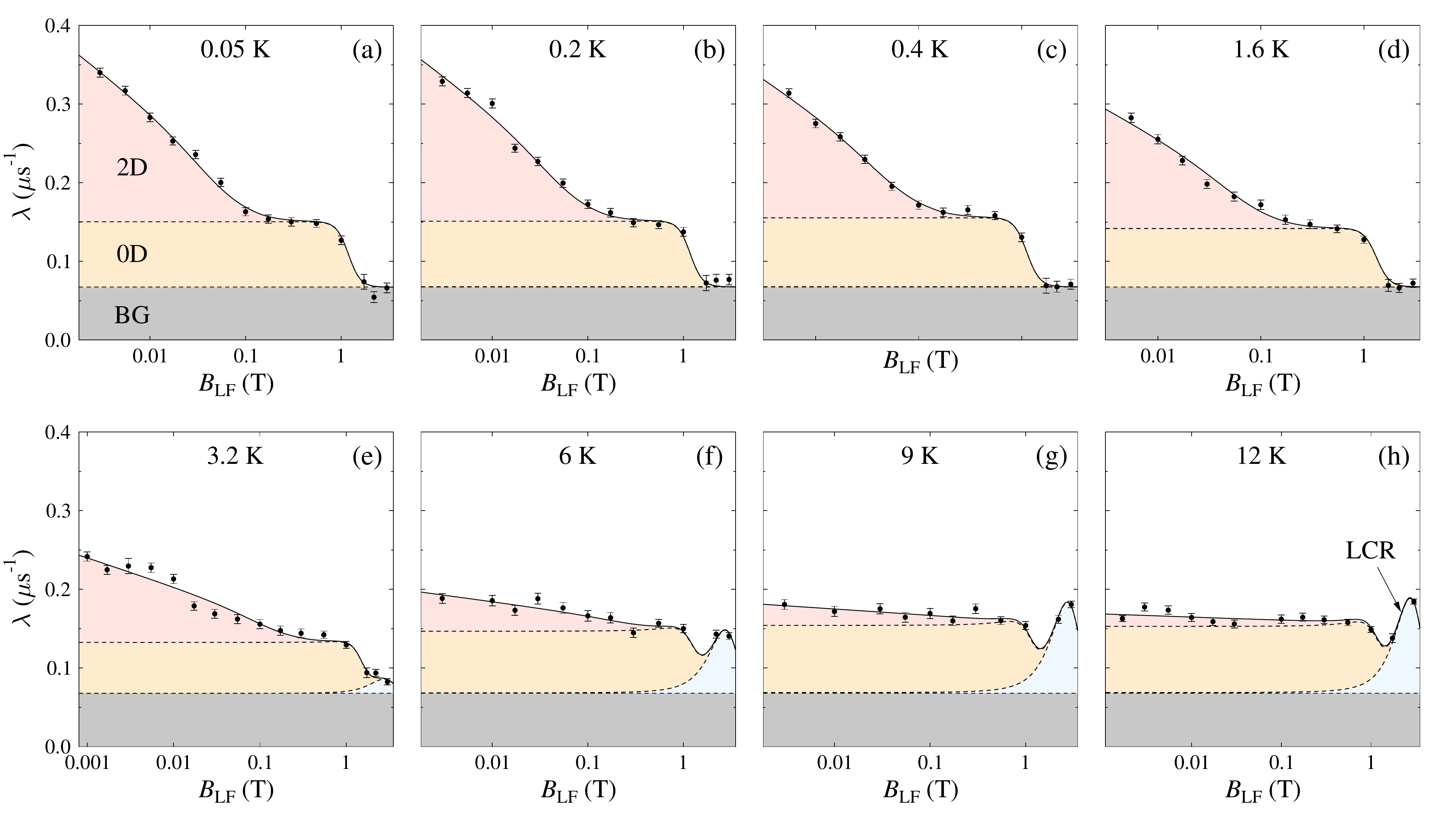}
\caption{\footnotesize 
LF-dependent relaxation rate measured at eight different temperatures.
The observed relaxation rate is resolved into four distinct spectral contributions distinguished by the shading [also see Eq.~(\ref{eqn1})]. The solid lines are the result of a combined global fitting to the field dependence.
The level-crossing resonance (LCR) is discussed in the End Matter.
}
\label{fig3}
\end{figure*}

\begin{figure}[htb]
\includegraphics[width = 0.4\textwidth]{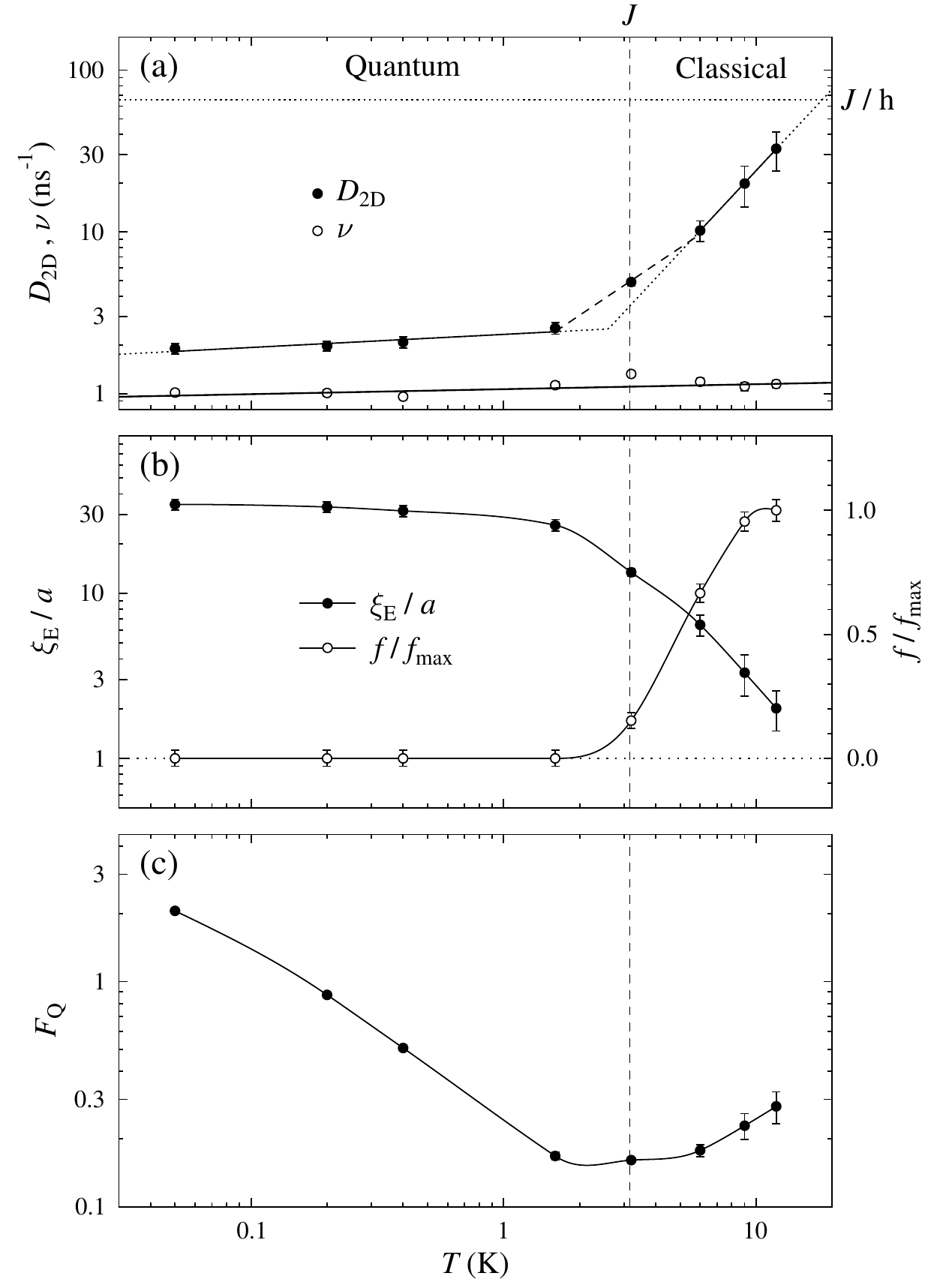}
\caption{\footnotesize 
Temperature-dependent fit parameters derived from a global fit of the LF-dependent data, showing a crossover between the quantum and classical regions.
(a) 2D diffusion rate and localized fluctuation rate.
(b) Entanglement length estimated from the diffusion rate (left hand scale) and amplitude of the high-field resonance (right hand scale).
(c) Quantum Fisher Information parameter derived from the fitted 2D term.
}
\label{fig4}
\end{figure}

Detailed LF scans were made at eight temperatures and the results are plotted in Fig.~\ref{fig3}.
The LF dependence is described by the sum of four terms
\begin{equation}
 \lambda(B_\mathrm{LF})  =  \lambda_\mathrm{2D}(B_\mathrm{LF})  +  \lambda_\mathrm{0D}(B_\mathrm{LF}) +  \lambda_\mathrm{BG} + \lambda_\mathrm{LCR}(B_\mathrm{LF}),\\
 \label{eqn1}
\end{equation}
with 
\begin{eqnarray}
\lambda_\mathrm{2D}(B_\mathrm{LF}) & = &  \frac{A^2}{4}  J_\mathrm{2D}(D_\mathrm{2D},\omega_e), \label{eqn2}\\
\label{eqn3}
\lambda_\mathrm{0D}(B_\mathrm{LF})  & = &  \frac{D^2}{4}  \frac{2/\nu} {1+(\omega_\mu/\nu)^m},\\
\label{eqn4}
\lambda_\mathrm{LCR}(B_\mathrm{LF})  & = & f~G(B_\mathrm{LF},B_\mathrm{0},B_\mathrm{wid}),
\end{eqnarray}
where $\lambda_\mathrm{BG}$ is the constant background relaxation rate, $J_\mathrm{2D}$ is the spectral density associated with two-dimensional (2D) diffusive spin fluctuations~\cite{Pratt2022},  $A$ and $D$ are hyperfine and dipolar coupling constants, $\omega_\mathrm{e} = \gamma_\mathrm{e} B_\mathrm{LF}$, $\omega_\mu = \gamma_\mu B_\mathrm{LF}$, and $G(B_\mathrm{LF},B_\mathrm{0},B_\mathrm{wid})$ is a Gaussian centred on $B_\mathrm{0}$ with width $B_\mathrm{wid}$. 
In fitting Eq.~(\ref{eqn2})--(\ref{eqn4}) to the eight LF scans, 
$D_\mathrm{2D}$, $\nu$ and $f$ were allowed to vary with $T$, whereas all other parameters were assumed independent of $T$ and estimated globally.
The globally determined parameters are listed in Table~\ref{tab:LFfits}. 
The $T$-dependent parameters $D_\mathrm{2D}$ and $\nu$ are plotted in Fig.~\ref{fig4}(a) and $f$ is shown in Fig.~\ref{fig4}(b) (right-hand axis). 
Following the procedure used in Ref.~\cite{Pratt2022}, we derived the $T$-dependent quantum entanglement length $\xi_\mathrm{E}$ [Fig.~\ref{fig4}(b)], which demonstrates that the spins become significantly entangled as $D_{\rm 2D}$ falls at low temperature, and the Quantum Fisher Information parameter $F_\mathrm{Q}$ [Fig.~\ref{fig4}(c)], which quantifies the development of multipartite entanglement associated with many-body correlations \cite{Hauke2016,ScheiePRB2021,Laurell2021} and here reveals a substantial increase on cooling well below $T\sim J$.

\renewcommand{\arraystretch}{1.3}
\vspace{0pt}
\begin{table}[htb]
	\caption{\label{tab:LFfits}%
		Global fit parameters defined in Eq.~(\ref{eqn1}) and obtained from the multi-temperature global fitting of the LF relaxation rate (Fig.~\ref{fig3}).}
	\begin{ruledtabular}
		\begin{tabular}{cccccc}
			$A$  & $D$ & $\lambda_\mathrm{BG}$ & $m$ & $B_0$ & $B_\mathrm{wid}$ \\
            (MHz) & (MHz) & (MHz) &  & (T) & (T) \\ 
			\colrule
			63(2) & 18.4(5) & 0.067(3)  & 7(1) & 2.7(1) & 1.3(1) \\
		\end{tabular}
	\end{ruledtabular}
\end{table}

\vspace{0pt}
\begin{table}[htb]
	\caption{\label{tab:D2Dfit}%
		Exchange coupling parameter $J$ estimated from (i) $\chi$ \cite{Bag2024} and from (ii) the  $D_\mathrm{2D}$ crossover [Fig.~\ref{fig4}(a)]. The final column gives the estimated exponent for the weak $T$-dependence  $D_\mathrm{2D} \propto T^{n_{\rm D}}$ observed in the $T<J$ regime.
	}
	\begin{ruledtabular}
		\begin{tabular}{ccccc}
			$J^{\rm (i)}$ & $J^{\rm (i)}/h$  & $J^{\rm (ii)}$ &  $J^{\rm (ii)}/h$ & $n_\mathrm{D}$ \\
             (K) & (ns$^{-1}$) & (K) & (ns$^{-1}$) &  \\ 
			\colrule
			3.16 & 66 & 2.6(8) & 54(17) & 0.08(3) \\
		\end{tabular}
	\end{ruledtabular}
\end{table}

The present results can be compared to those obtained previously for YbZnGaO$_4$~\cite{Pratt2022} in which a similar crossover was observed between classical and quantum entangled regions, centered around the slightly higher $J$ value. Here, the $J$ value obtained from the extrapolated intersection of the low-$T$ and high-$T$ dependences of $D_\mathrm{2D}$ [Fig.~\ref{fig4}(a)] is consistent with the more accurate value obtained from the magnetic susceptibility (Table~\ref{tab:D2Dfit}).
One significant enhancement over the earlier study in Ref.~\onlinecite{Pratt2022} is the extended field range of the HiFi instrument used here, which has allowed us to measure a particularly sharp cutoff in part of the non-2D contribution, where $m$ = 7(1) was obtained from the global fit, as well as revealing a high-field resonance around 3~T for temperatures larger than $J$ (see Fig.~\ref{fig3}(e) to Fig.~\ref{fig3}(h)). The size of the entanglement parameters reached in the quantum region for YbZn$_2$GaO$_5$ are both larger than found previously for YbZnGaO$_4$, which may reflect the absence of non-magnetic site disorder in the present system.

\begin{table}[b]
	\caption{\label{tab:sites}%
		Muon stopping sites in YbZn$_2$GaO$_5$ obtained from DFT using the structural relaxation method, where sites A to C are visualised in Fig.~\ref{fig1}(a). For each site, we show the energy $E$ relative to the lowest energy site, estimates for the relaxation $\Delta$ due to nuclear magnetic fields, and the occurrence of each site based on 30 relaxation calculations.
	}
	\begin{ruledtabular}
		\begin{tabular}{lcccr}
			\textrm{Site}&
			\textrm{Fractional }&
			$E$ & 
            $\Delta$  & 
            Number of \\
            & \textrm{coordinates}
            & (eV)
            & (MHz)
            & structures \\
			\colrule
			A & (0.536 0.072 0.068) & 0.00 &  0.025 & 13 \\
			B & (0.813 0.906 0.112) & 0.24 & 0.033 & 9 \\
			C & (0.260 0.747 0.206) & 0.92  & 0.154 & 6 \\
            D & (0.667 0.333 0.504) & 1.31  & 0.038 & 2 \\
		\end{tabular}
	\end{ruledtabular}
\end{table}

\begin{table}[htb]
	\caption{\label{tab:sitecouplings}%
		Dipolar coupling parameters between the calculated muon sites and their closest Yb$^{3+}$ ions, with the Yb moments taken to be 2.1(1) $\mu_\mathrm{B}$. The final column indicates whether a significant contact hyperfine term is expected in addition to the dipolar term.
	}
	\begin{ruledtabular}
		\begin{tabular}{lcccc}
			Muon site &
			Yb label &
			Yb distance &
            $D$ &
            $A$ term \\
             & & (\AA) & (MHz) & \\
			\colrule
			A & 1a,1b & 2.446 & 36(2) & \checkmark  \\
			 & 2 & 3.102 & 17.7(8) &   \\
			 & 3 & 3.396 & 13.5(6) &  \checkmark \\
			\colrule
			B & 1 & 2.650 & 28(1) &   \\
			 & 2a,2b & 3.741 & 10.1(5) &   \\
			 & 3a,3b & 4.258 & 6.8(3) &   \\
			 & 4a,4b & 4.648 & 5.3(3) &   \\
			\colrule
			C & 1 & 4.861 & 4.6(2) &   \\
			 & 2a,2b & 5.120 & 3.9(2) &   \\
		\end{tabular}
	\end{ruledtabular}
\end{table}

Using DFT (in particular, the DFT+$\mu$ method~\cite{Moller2013_FuF,Bernardini2013,Blundell2023}), it is possible to identify the muon site. Here, we have carried out density functional theory (DFT) calculations using the plane-wave basis-set electronic structure code \textsc{castep}~\cite{CASTEP} (for details, see End Matter). 
Our calculations reveal four crystallographically distinct muon stopping sites, whose properties are summarized in Table~\ref{tab:sites}. In each case, the muon forms a bond with an oxygen atom, with a $\mu^+$--O bond length of $\approx 1$~\AA{}. However, the oxygen involved is different in each case. For site A, as shown in Fig.~\ref{fig1}(a), the muon is bonded to an oxygen which is bonded to Yb (O1); for site B, it is bonded to an oxygen bonded to Zn and Ga (O3); and for site C, it is bonded to an oxygen bonded to Ga only (O2). We also find an additional site, labelled D, where the muon sits in the center of a Yb triangle, forming a bond O1 that lies along the $c$ axis. However, this site is very high in energy, and is only found twice among our relaxed structures, suggesting a small basin of attraction, so unlikely to be realized in practice. We found the effect of the muon on local crystal fields to be unimportant (see End Matter), as expected for symmetry-protected Kramers doublets.

We have estimated the relaxation rate, $\Delta$, due to nuclear field from the van Vleck second moment in the limit of quadrupolar coupling~\cite{Hayano1979}. The nuclear relaxation rate is dominated by the contributions of Ga, which exists as two stable $I=3/2$ isotopes. Hence, muons at site C, sitting only $\approx 2$~\AA{} away from the nearest Ga experience a much larger nuclear relaxation than muons sitting at either site A or site B. However, these nuclear fields are quenched upon the application of a sufficiently large longitudinal field, leaving the muons instead sensitive to the fields due to fluctuating electronic moments.

The dipolar coupling parameters between the muon and the closest Yb$^{3+}$ ions have been evaluated for these sites, with the results shown in Table~\ref{tab:sitecouplings}. 
Our $\mu$SR results can be fully explained by assuming that only the lowest energy A site is present. 
The $\lambda_\text{2D}$ term in Eq.~(\ref{eqn1}) is then assigned to interaction with Yb3, which is expected to have the largest contact hyperfine coupling, taking place across a bridging oxygen. 
The $\lambda_\text{0D}$ term is assigned to interaction with Yb2, which is expected to only have dipolar coupling. The calculated dipolar parameter for Yb2 is consistent with the fitted $D$ value for the 0D term.  
The pure dipolar coupling of Yb2 also allows it to produce the high-field level crossing resonance. 
The field-independent part of the $\lambda_\text{BG}$ term may be assigned to interaction with Yb1a and Yb1b, which are the closest ions to the muon. The combination of fast fluctuations and strong coupling for these ions could lead to the observed field-independent response in the experimentally accessible field range.

In contrast to classically ordered states, a
QSL does not break symmetry.  A QSL can be classified by its projective symmetry group (PSG), formed by the transformations that keep the mean-field description unchanged.  These transformations are composed of a combination of symmetry operations (e.g.\ translations or rotations) and gauge transformations \cite{Wen2002}.  For Yb$^{3+}$-containing compounds, the spin-orbital entanglement results in additional features of symmetry operations acting on the lattice, resulting in a rich structure of possible PSGs \cite{LiWangChen2016,Li2017}.  Our results are consistent with U1A01 QSL in the classification scheme of Ref.~\onlinecite{Li2017}, the `U1' denoting a $U(1)$ spin liquid, and the `A' denoting zero (rather than $\pi$) flux per triangular plaquette.  This theory has a linear (Dirac) dispersion, consistent with our measured small spin-diffusion power law (and with the quadratic power law for the heat capacity \cite{Bag2024}).
This is in contrast to the more complicated U1A11 state, with both linear and quadratic dispersion, that had to be invoked previously to understand the properties of YbZnGaO$_4$~\cite{Pratt2022}.  Our results show how detailed measurements of low-freqency spin dynamics in a chemically-ordered rare earth triangular magnet can not only demonstrate spin liquid behavior but also substantially constrain the type of topological order present in the ground state.

{\it Acknowledgments:}
The $\mu$SR experiments were carried out at the ISIS neutron and muon source, UK, and were supported by beam-time allocation RB2310372. We acknowledge support by the ISIS Neutron and Muon Source, the scholarship funding by the Croucher Foundation, and UK Research and Innovation (UKRI) under the UK government’s Horizon Europe funding guarantee [Grant No. EP/X025861/1]. The pulsed magnet system as the Nicholas Kurti High Magnetic Field Laboratory was refurbished 
with a grant from
the UK Engineering and Physical Sciences Research Council (EPSRC)
[Grant No.\ EP/J013501/1].
D.\,P. acknowledges the financial support by the Oxford-ShanghaiTech Collaboration Project and EPSRC [Grant No.\ EP/T028637/1]. J.\,S. acknowledges support from National Science Foundation Cooperative Agreement No. DMR-2128556 and the US Department of Energy Basic Energy Science program “Science at 100 T”. J.\,S. is also grateful to the University of Oxford for the provision of a Visiting Professorship that permitted some of the measurements described in this paper.

\bibliography{library.bib}

\onecolumngrid
\section*{End matter}
\twocolumngrid

\begin{figure}[htb]
\includegraphics[width = 0.48\textwidth]{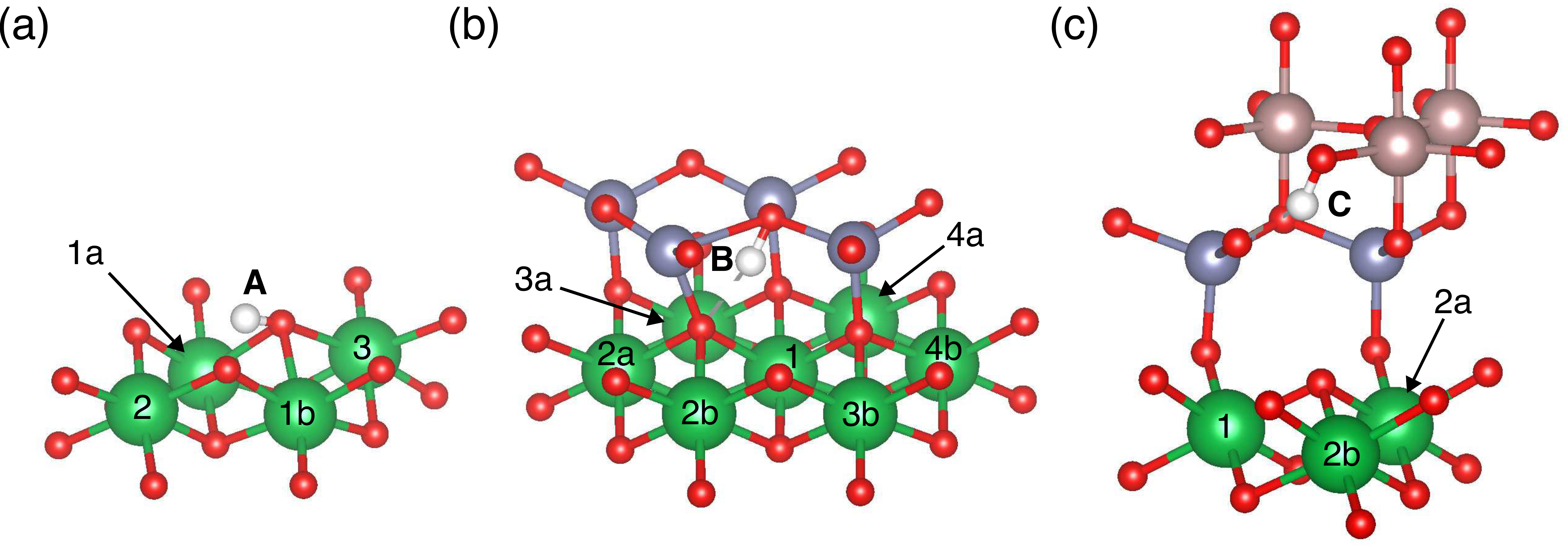}
\caption{\footnotesize 
Local geometry around the muon site for (a) site A, (b) site B, and (c) site C. Nearby Yb$^{3+}$ ions are labelled according to the scheme in Table~\ref{tab:sitecouplings}.}
\label{fig:local_geometry}
\end{figure}

\begin{figure}[htb]
\includegraphics[width = 0.48\textwidth]{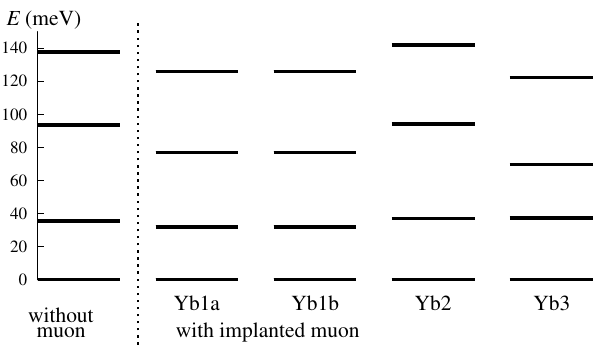}
\caption{\footnotesize 
Calculated CF levels of Yb$^{3+}$ ions with and
without an implanted muon and its accompanying lattice distortion in YbZn$_2$GaO$_5$. Energy levels
are calculated for the four distinct Yb$^{3+}$ ions nearest from the muon site A. Each line indicates a doublet state.}
\label{fig:CF_level}
\end{figure}

\begin{figure}[htb]
\includegraphics[width = 0.33\textwidth]{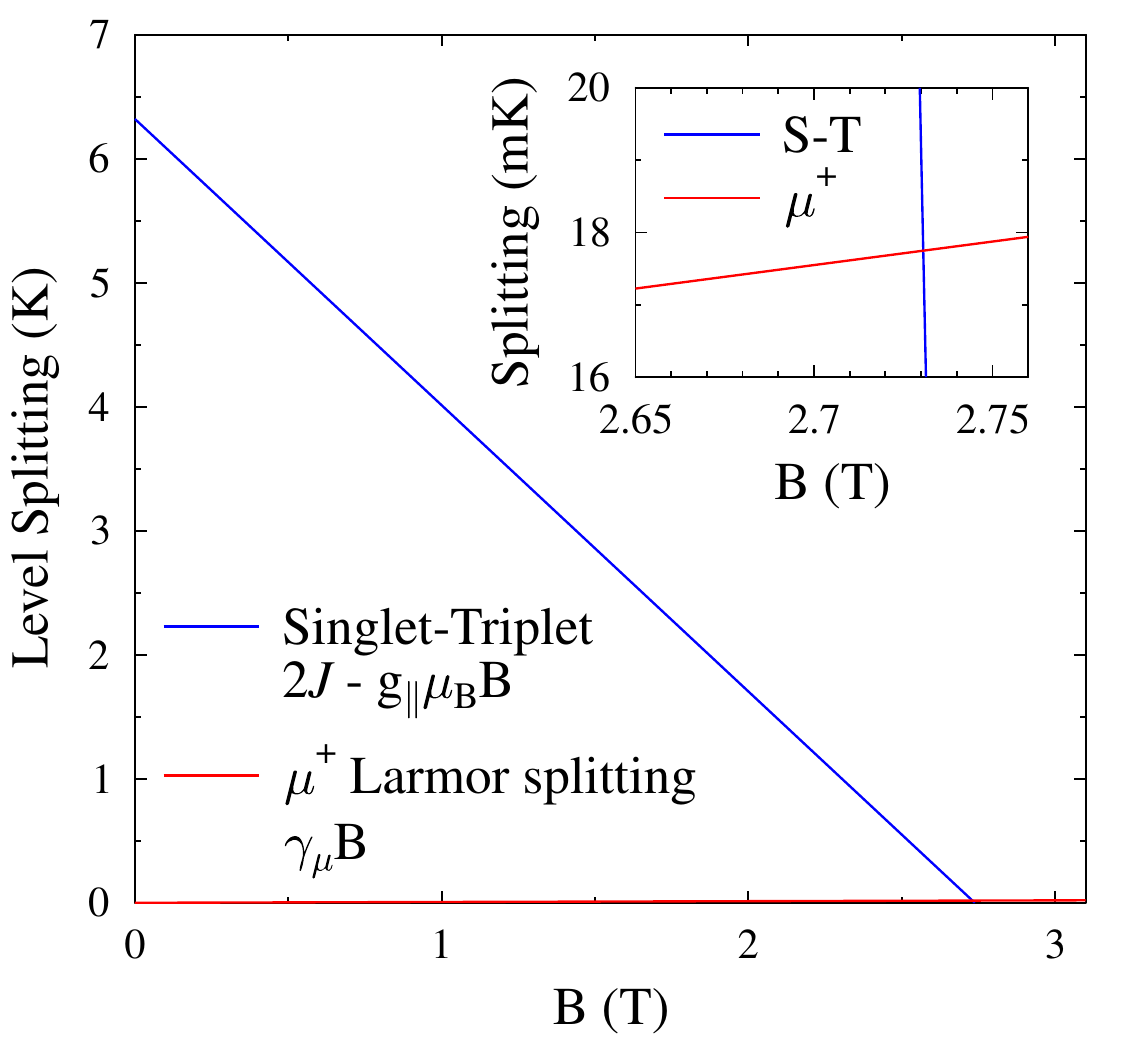}
\caption{\footnotesize 
Origin of the 2.7~T level crossing resonance observed in the data at higher $T$.
The singlet-triplet splitting of a pair of Yb$^{3+}$ ions is linearly reduced by magnetic field. Taking the reported values of $J$ = 3.16~K and $g_\parallel$ = 3.436, the singlet-triplet gap will match the muon Larmor frequency at a field of 2.73~T, leading to a level-crossing resonance between the muon and the pair of Yb$^{3+}$ spins. 
The inset expands the level crossing region.}
\label{fig5}
\end{figure}

{\it Pulsed-field magnetometry:}
A sample of YbZn$_2$GaO$_5$ was loaded into an ampule and affixed to the end of a probe. The probe was placed in a $^3$He cryostat where a temperature of 0.45~K could be achieved. The sample could be inserted into and extracted from a compensated coil, in situ, enabling measurement of the sample’s magnetization. Measurements were made using a a 50~T pulsed magnet ($\sim$10~ms pulse width) at the Nicholas Kurti Magnetic Field Laboratory, Oxford University, UK.

{\it DFT$+\mu$ calculations:}
Calculations were carried out within the generalized-gradient approximation (GGA) using the PBE functional~\cite{PBE} and the system was treated as non-spin-polarized. For our calculations, we use a plane-wave cutoff energy of 900 eV and $5\times 5 \times 2$ Monkhorst-Pack grid~\cite{mpgrid1976} for Brillouin zone sampling, resulting in total energies that converge to 2 meV per atom and forces that converge to $5\times 10^{-2}$ eV \AA$^{-1}$. Muon site calculations were performed using the MuFinder program~\cite{mufinder}. Initial structures comprising a muon and a $2 \times 2 \times 1$ supercell of YbZn$_2$GaO$_5$ were generated by requiring the muon to be at least 0.5~\AA~away from each of the muons in the previously-generated structures (including their symmetry equivalent positions) and at least 1.0~\AA~away from any of the atoms in the cell.  This resulted in 52 structures that were subsequently allowed to relax until the calculated forces on the atoms were all $<5\times 10^{-2}$ eV \AA$^{-1}$ and the total energy
and atomic positions converged to within $2\times10^{-5}$ eV per atom and $1\times10^{-3}$~\AA, respectively. Some initial positions resulted in geometries that we were unable to converge. These were discarded, leaving us with 30 relaxed supercells.

{\it Crystal field calculations:}
We also quantify the effect of muon-induced distortions on the crystal field (CF) levels of the nearby Yb$^{3+}$ ions. Using a point-charge model generated using the PyCrystalField library~\cite{Scheie2021}, we compare the CF levels in the unperturbed structure with those of each of the Yb$^{3+}$ in the layer containing the implanted muon. We also include the muon as an additional point charge coordinating the Yb$^{3+}$. The resulting energy levels are shown in Fig.~\ref{fig:CF_level}. The calculated gap between the ground state and first-excited state without the muon is in good agreement with the value determined from experiment \cite{Bag2024}.  With the implanted muon, this gap remains in the range 30--40~meV, which is broadly consistent with the activation energy $E_{\rm A}=25(2)$~meV extracted from the fit of $\lambda(T)$ given earlier. The CF levels for Yb$1$a, Yb$1$b and Yb$3$ are very similar, with each of these being slightly lowered compared to the corresponding level in the undistorted case. Yb$2$ is distinct in that the CF levels are instead slightly raised. However, the effect of the distortion field on the CF levels of this Kramers ion is calculated to be rather small, in contrast to the case of Pr-based pyrochlores~\cite{Foronda2015} (Pr$^{3+}$ is a non-Kramers ion and its energy levels are consequently vastly more susceptible to the effects of a muon-induced distortion of local ions). Hence, we do not expect the presence of the muon to modify the local magnetic properties to which the muon is sensitive.

{\it Level-crossing resonance:}
The high-field feature at 2.7~T in Fig.~\ref{fig3} appears to be a new type of muon level crossing resonance.  This interpretation is illustrated in Fig.~\ref{fig5}, where the calculated field dependence of the singlet-triplet splitting of an interacting pair of Yb$^{3+}$ ions matches the Larmor splitting of a diamagnetic muon at a predicted field of 2.73~T, fully consistent with the experimental value of 2.7(1)~T. It is very interesting to see that this resonance only appears in the high-$T$ paramagnetic state and becomes strongly suppressed in the quantum entangled low-$T$ state. 

{\it Quantum Fisher information:}  A state is entangled if its density matrix is not separable.  An entanglement witness is a functional of the density matrix that can be used to distinguish between an entangled state and a non-entangled (separable) state.  The Quantum Fisher information $F_{\rm Q}$ has been shown to act as a witness of multipartite entanglement \cite{Hyllus2012,Toth2012}, and can be expressed in terms of the imaginary part of the dynamical susceptibility $\chi''(\omega)$ as
$F_{\rm Q} = {4\over\pi}\int_0^\infty \tanh \left( {\hbar\omega\over 2 k_{\rm B}T} \right) \chi''(\omega,T)\,{\rm d}\omega$ \cite{Hauke2016}.  Using the fluctuation-dissipation theorem, we can express this in terms of the spectral density function measured by the muon using
$F_{\rm Q} = {4\over\pi}\int_0^\infty \tanh^2 \left( {\hbar\omega\over 2 k_{\rm B}T} \right) J_{\rm 2D}(\omega)\,{\rm d}\omega$.

\end{document}